\documentclass[showpacs,amsmath,twocolumn,showkeys,pra,superscriptaddress]{revtex4-1}
\usepackage{dcolumn}    % Align table columns on decimal point
\usepackage{ifpdf}
\usepackage{amssymb,lineno,amsfonts}
\usepackage{graphicx}   % Include figure files
\usepackage{dcolumn}    % Align table columns on decimal point
\usepackage{bm}         % bold math
\usepackage{bbm}
\usepackage{mathrsfs}
\usepackage{upgreek}
\usepackage{mathtools}
\usepackage{epstopdf}
\usepackage{setspace}
\usepackage{hyperref}
\usepackage{float}
\usepackage{natbib}
\usepackage[usenames,dvipsnames]{xcolor}

\newcommand{\ket}[1]{\vert{#1}\rangle}

\newcommand{\outpr}[2]{\vert{#1}\rangle\langle{#2}\vert}

\newcommand{\expec}[1]{\langle{#1}\rangle}

\definecolor{med-blue}{RGB}{25,25,112}
\hypersetup{colorlinks, linkcolor={red},citecolor={blue}, urlcolor={MidnightBlue}}

\begin{document}
\title{NMR investigation of contextuality in a quantum harmonic oscillator \\
via pseudospin mapping}
\author{Hemant Katiyar} 
\email{hkatiyar@uwaterloo.ca}
\affiliation{NMR Research Center, Indian Institute of Science Education and Research, Pune 411008, India} 
\author{C. S. Sudheer Kumar} 
\email{sudheer.kumar@students.iiserpune.ac.in}
\affiliation{NMR Research Center, Indian Institute of Science Education and Research, Pune 411008, India} 
\author{T. S. Mahesh} 
\email{mahesh.ts@iiserpune.ac.in}
\affiliation{NMR Research Center, Indian Institute of Science Education and Research, Pune 411008, India} 

\begin{abstract}
Physical potentials are routinely approximated to harmonic potentials so as to analytically solve the system dynamics.  Often it is important to know when a 
quantum harmonic oscillator (QHO) behaves quantum mechanically and when classically.
Recently Su et. al. [Phys. Rev. A {\bf 85}, 052126 (2012)] have theoretically shown that
QHO exhibits quantum contextuality (QC) for a certain set of pseudospin observables.  
In this work, we encode the four eigenstates of a QHO onto four Zeeman product states of a pair of spin-1/2 nuclei. Using the techniques of NMR quantum information processing, we then demonstrate the violation of a state-dependent inequality
arising from the noncontextual hidden variable model, under specific experimental arrangements.  We also experimentally demonstrate the violation of a state-independent inequality by thermal equilibrium states of nuclear spins, thereby assessing their quantumness.
\end{abstract}

%\pacs{03.67.Lx, 03.65.Ta, 03.67.-a, 03.65.-w, 82.56.-b}
\maketitle

\section{Introduction}
Quantum contextuality (QC) states that the outcome of the measurement depends
not only on the system and the observable but also on the 
context of the measurement, i.e., on other compatible observables which are measured along with \cite{peres_context_1pg,Peres,quant_theory_peres,KS}.

Let us consider a pair of space-like separated entangled particles, with local observables $A$ and $C$ belonging to the first particle, and $B$ and $D$ to the second.  We assume that these observables are dichotomic (i.e., can take values $\pm 1$) and that the pairs $ (A,B) , ~ (B,C), ~ (C,D), $ and $(D,A)$ commute.

Classically, one assigns objective properties to the particles such that $D$ behaves identically on the state of the system irrespective of whether it is measured in the context of $A$ or in the context of $C$, even though $A$ and $C$ are not compatible \cite{Mermin,EPR_original_paper}. Such measurements are said to be \textit{context independent}.  Classically, one can pre-assign values  $(a,c)$ to $(A,C)$ of the first particle independent of the measurement carried out on the second particle.  Similarly, for the second particle one can pre-assign values $(b,d)$ to $(B,D)$ independent of the measurement carried out on the first particle. In these pre-assignments, implicit is the assumption of noncontextual hidden variables, which predict definite measurement outcomes independent of measuring arrangement. 
If we pre-assign values to observables such that $A,B,C,D = \pm 1$, it follows that $AB+BC+CD-AD=\pm 2$ and the expectation value,
\begin{eqnarray}
\mathrm{\textbf{I}}~ && =\expec{AB+BC+CD-AD} \nonumber \\
&& = \expec{AB} + \expec{BC} + \expec{CD} - \expec{AD} \leq 2
\label{I_main}
\end{eqnarray}
\cite{quant_info_neilson_chuang}.  This inequality often known as 
CHSH inequality arises from noncontextual hidden variable (NCHV) model
and must be satisfied by all classical particles.

Now let us see the implication of the quantum theory. Let Alice and Bob share a large number of singlet states: $(\ket{01}-\ket{10})/\sqrt{2}=-(\ket{+-}-\ket{-+})/\sqrt{2}$, where $\ket{0}$ and $\ket{1}$ are eigenkets of Pauli-$z$ operator ($\sigma_z$) and $\ket{\pm}=(\ket{0}\pm\ket{1})/\sqrt{2}$. Alice measures on her qubit either $\sigma_x^A$ or $\sigma_z^A$, while Bob always measures $\sigma_x^B$. Let us compare the results of only those measurements in which Alice has obtained outcome $+1$.
If Alice measures $\sigma_z^A$, then Bob's qubit collapse to 
$\ket{1}=(\ket{+}-\ket{-})/\sqrt{2}$. In this context (i.e., $\sigma_z^A$), Bob will get both outcomes $\pm 1$ with equal probability. 
On the other hand, if Alice measures $\sigma_x^A$ on her qubit, then Bob's qubit collapses to $\ket{-}$ and in this context (i.e., $\sigma_x^A$), Bob will always get the outcome $-1$.  Hence the context dependency.

Here we experimentally investigate QC of a quantum harmonic
oscillator (QHO). There are a variety of quantum systems whose potentials
are approximated by QHO.
Consider for example the quantized electromagnetic field used to manipulate a qubit in cavity quantum electrodynamics\cite{cQED_review}.
Recently, QC in QHO has been theoretically studied by Su et al \cite{Cont_theory}
by mapping four lowermost QHO states onto four pseudospin states.  Such
states can be encoded by qubit states, and QC can be studied by
realizing the measurements of appropriate observables.  
We realize this study using a nuclear magnetic resonance (NMR) quantum simulator \cite{Corynmr_1st}
.

In the following section we shall revisit the formulation of Su et al, and in section III, we describe the experimental demonstration of state-dependent and
state-independent QC using an NMR system.  Finally we conclude in section IV.

\section{theory}
Hong-Yi Su \textit{et.al.} \cite{Cont_theory} have theoretically studied QC of eigenstates of $1$D-QHO by introducing
two sets of pseudo-spin operators,
\begin{eqnarray}
{\bf \Gamma } = (\Gamma_x,\Gamma_y,\Gamma_z),~~
{\bf \Gamma '} = (\Gamma_x',\Gamma_y',\Gamma_z') \nonumber
\end{eqnarray}
with components,
\begin{eqnarray}
\Gamma_x = \sigma_x \otimes \mathbbm{1},
\Gamma_y = \sigma_z \otimes \sigma_y,
\Gamma_z = -\sigma_y \otimes \sigma_y,  \nonumber \\
\Gamma_x' = \sigma_x \otimes \sigma_z,
\Gamma_y' = \mathbbm{1} \otimes \sigma_y,
\Gamma_z' = -\sigma_x \otimes \sigma_x.  
\label{Gammas defined}
\end{eqnarray}
where $ \mathbbm{1} $ is $2\times 2$ identity matrix. Using these operators they defined the following observables,
\begin{eqnarray}
A&=&\Gamma_x = \sigma_x \otimes \mathbbm{1}, \nonumber \\
B&=&\Gamma_x' \cos \beta  + \Gamma_z' \sin \beta =\sigma_x\otimes(\sigma_z\cos \beta  - \sigma_x\sin \beta ),  \nonumber \\
C&=&\Gamma_z = -\sigma_y \otimes \sigma_y, \nonumber \\
D&=&\Gamma_x'\cos \eta  +  \Gamma_z'\sin\eta=\sigma_x\otimes(\sigma_z\cos \eta  - \sigma_x\sin \eta ). 
\label{A,B,C,D defined}
\end{eqnarray}
The products which form the inequality expression (Eq. \ref{I_main}) are
\begin{eqnarray}
 AB &=&\mathbbm{1}\otimes(\cos\beta~\sigma_z-\sin\beta~\sigma_x),\nonumber\\
 BC &=&-\sigma_z\otimes(\cos\beta~\sigma_x+\sin\beta~\sigma_z)\nonumber\\
 CD &=&-\sigma_z\otimes(\cos\eta~\sigma_x+\sin\eta~\sigma_z),\nonumber\\
 DA &=& \mathbbm{1}\otimes(\cos\eta~\sigma_z-\sin\eta~\sigma_x).
 \label{AB,BC,..evalted}
\end{eqnarray}
Following commutation relations hold: $ [\Gamma_i,\Gamma_j'] = 0~(i,j=x,y,z)$, $[\Gamma_x,\Gamma_y]=2i~\Gamma_z$ , $[\Gamma'_x,\Gamma'_y]=2i~\Gamma'_z$ and cyclic permutation of $x,y,z$. $ A,B,C,D $ have eigenvalues $ \pm 1 $, with $ (A,B)$, $(B,C)$, $(C,D)$ and $(D,A) $ forming compatible pairs. $A,~B,~C$ and $D$ are Hermitian, hence observables. One can verify that they are also unitary operators. Hong-Yi Su \textit{et.al.} \cite{Cont_theory} have shown that,
\begin{eqnarray}
\mathrm{\textbf{I}}_{\ket{l}_{QHO}}^{QM} = 2\sqrt{2} > 2 ,~\mathrm{when},~(\beta,\eta)_l = 
\begin{cases}
    (-\pi/4 , -3\pi/4)_0 \\
    (3\pi/4, \pi/4)_1 \\
    (\pi/4, 3\pi/4)_2 \\
    (-3\pi/4, -\pi/4)_3 
\end{cases} 
\label{max vio angles}
\end{eqnarray}
where, $\mathrm{\textbf{I}}_{\ket{l}_{QHO}}^{QM}$ is the expression on LHS of inequality \ref{I_main}, $ l=0,1,2\textrm{ and } 3 $, and, $ \ket{0}_{QHO}, \ket{1}_{QHO}, \ket{2}_{QHO}$ and $ \ket{3}_{QHO}$ are first four energy eigenstates of 1D-QHO.  Thus QHO violates the inequality (\ref{I_main}) for certain
observables and thereby exhibits QC.

It is well known that only certain two-particle states violate the CHSH inequality (\ref{I_main}). As shown in \cite{D_Home_book,capasso} factorable states always satisfy inequality (\ref{I_main}) for local observables (observables of the form $P \otimes \mathbbm{1}$ or $\mathbbm{1} \otimes Q$ \cite{Audruch_entangled_sys}). With maximally mixed state ($\mathbbm{1}\otimes\mathbbm{1}/4$) the inequality (\ref{I_main}) is satisfied even with nonlocal observables (observables of the form $P \otimes Q$ \cite{Audruch_entangled_sys}) in eq. (\ref{A,B,C,D defined}), which is obvious from the fact that all the products in
eq. (\ref{AB,BC,..evalted}) are traceless. However, if the initial state is nonfactorable, we can always find observables such that inequality (\ref{I_main}) is violated \cite{D_Home_book}.  Although the pseudospin states $\{\ket{00},\ket{01},\ket{10},\ket{11}\}$, are factorable, they still violate (\ref{I_main}) since the observables in eq. (\ref{A,B,C,D defined}) are nonlocal.  
Thus, we observe that even when a system is in a nonentangled state, 
measurements of nonlocal observables may lead to violation of noncontextuality 
inequality \cite{Cabello_NCHV_ineqlty}.

\textbf{State independent QC:} 
There exist stronger inequalities obtained from NCHV models which is violated by
all states, including separable or maximally mixed states.
 If the initial state is maximally mixed, entanglement cannot be created by measuring whatever observable (local or nonlocal). This shows that entanglement is not necessary in general even in a bipartite system, to exhibit QC. Hence we conclude that, QC is more fundamental or general than entanglement. Any system whose Hilbert space has dimension $>2$ exhibits QC \cite{KS}. Even a \textit{single} spin-1 particle (where entanglement has no meaning as far as spin degree of freedom is concerned) can exhibit QC \cite{spin_1_QC}.

\section{Experiment}
\subsection{State dependent contextuality}
\begin{figure}
\centering
\includegraphics[width=7cm]{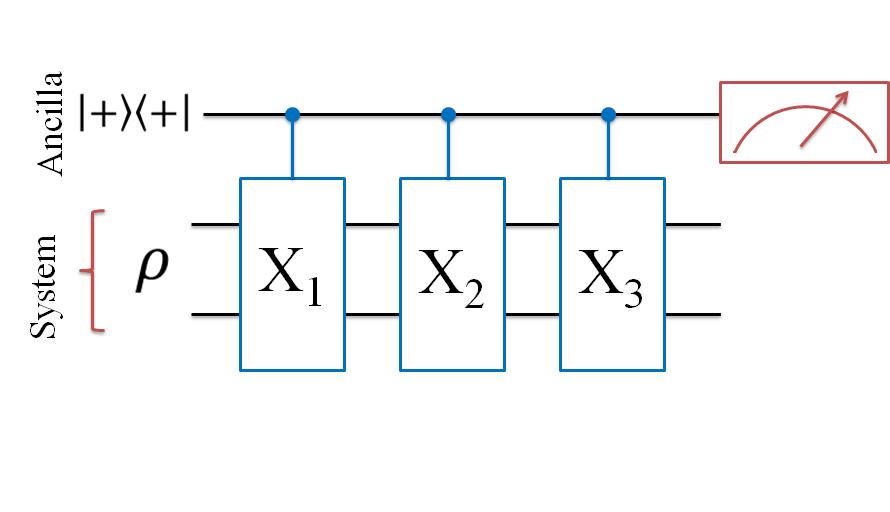}
\caption{Moussa Protocol for finding out the expectation value of the joint observable $ X_1X_2X_3 $ i.e. $ \expec{X_1X_2X_3} $. Here $X_i$'s are mutually commuting unitary operators.} 
\label{moussa}
\end{figure}

For experimentally studying eq. (\ref{I_main}), we need: (i) a physical representation of first four energy eigenstates $ \{ \ket{0}_{QHO}, \ket{1}_{QHO}, \ket{2}_{QHO}, \ket{3}_{QHO} \}$ of 1D-QHO , and (ii) a way to find out the expectation values for operators $ AB,~ BC, ~CD,$ and $DA $.
We encode the first four energy eigenstates of 1D-QHO onto the four energy eigenstates (under secular approximation) of a pair of spin-1/2 nuclei 
precessing in external static magnetic field: $\{\ket{00},\ket{01},\ket{10},\ket{11}\}$, which we call Zeeman product states. In fact any four arbitrarily choosen energy eigenstates of 1D-QHO and also their superposition states exhibit QC \cite{Cont_theory}. The circuit shown in Fig. \ref{moussa} is called Moussa protocol \cite{Moussa}, and is used to extract the expectation value of observables in a joint measurement.  This has subsequently been generalized by Joshi et al \cite{joshi_frankcond} to unitary operators.

\begin{figure}
\hspace*{-0.6cm}
\includegraphics[width= 9.5cm]{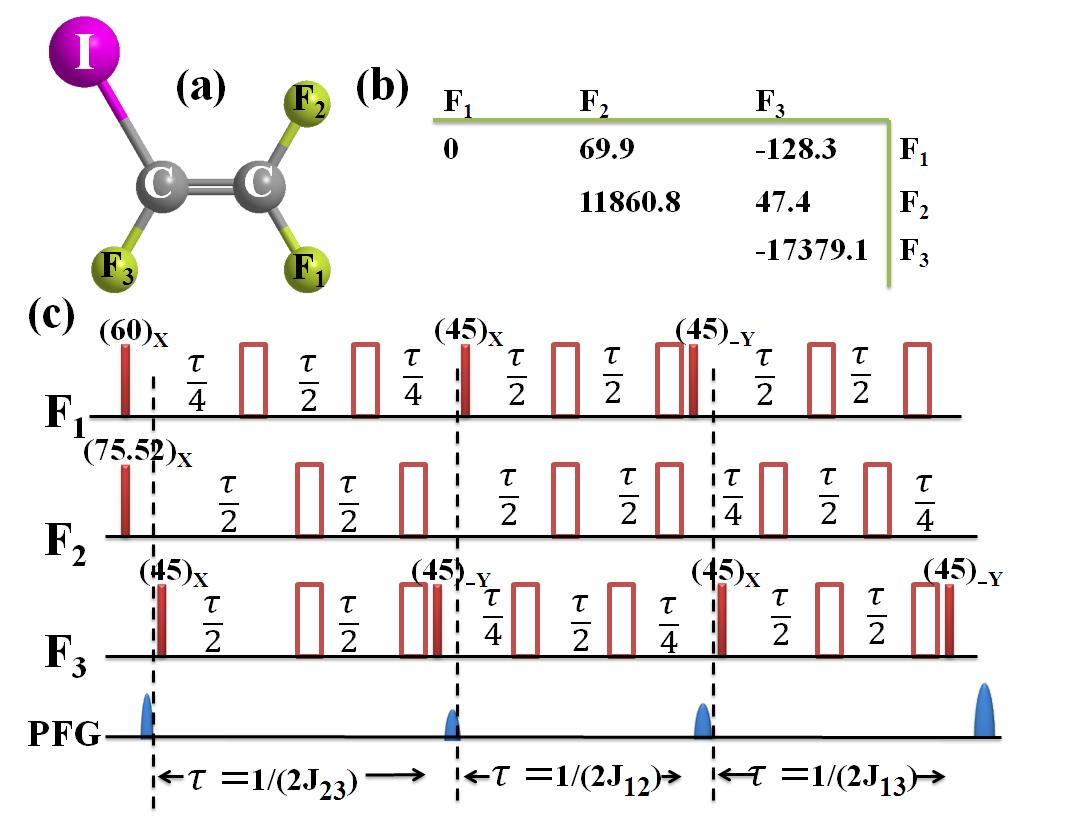}
\caption{(a) Molecular Structure, (b) chemical shifts(diagonal elements) and J-couplings(off-diagonal elements) in Hz of trifluoroiodoethylene, and (c) pulse sequence for pseudo-pure state preparation.
The amplitude and phase of the shaded pulse is written over the top of respective pulse, and unshaded pulses are $ (\pi)_x $ pulses. During the J evolution $1/(2J_{23})$, $ (\pi)_x $ pulses refocus $J_{12}$ and $J_{13}$ evolutions and retain only $J_{23}$ evolution. Similarly others. Last row represents Pulsed Field Gradients(PFG), which kills transverse magnetization or in other words they make coefficients of $ I_x  $ and $ I_y $ zero \cite{cavanagh}}
\label{molecule}
\end{figure}

The three qubits for this experiment were provided by the three $^{19}$F nuclear spins of trifluoroiodoethylene dissolved in acetone-D6. Fig. \ref{molecule}(a) shows the structure of trifluoroiodoethylene along with the Hamiltonian parameters in Fig. \ref{molecule}(b).
The effective $^{19}$F spin-spin (T$ _2^* $) and spin-lattice (T$ _1 $) relaxation time constants  were about $ 0.8 $ and $ 6.3 $ s respectively. The experiments were carried out at an ambient temperature of 290 K on a 500 MHz Bruker UltraShield NMR spectrometer.

The thermal equilibrium state for the three spin system in the eigenbasis of total Hamiltonian (under secular approximation: $\{\ket{000},\ket{001},...\}$) is \begin{equation}
\rho_{\mathrm{eq}} = \frac{\mathbbm{1}_8}{8} +\epsilon \sum_{i=1}^3 I_{iz}
\label{rho_eq}
\end{equation}
where, $\mathbbm{1}_8$ is an $8\times 8$ identity matrix, $I_{iz}$ are spin angular momentum operators, and the purity factor $\epsilon = \hbar \omega_0/(8kT)$ is the ratio of  the Zeeman energy gap to the thermal energy \cite{cavanagh}.
Unitary operation has no effect on the identity part, but
modifies only the traceless deviation part.
By applying a series of unitary and nonunitary operators (pulse sequence
shown in Fig. \ref{molecule} \cite{Avik}), it is
possible to transform the equilibrium state to a pseudopure state 
\begin{eqnarray}
\rho_\mathrm{pps}= (1-\epsilon)\frac{\mathbbm{1}_8}{8} + \epsilon \outpr{000}{000}
 =   \frac{\mathbbm{1}_8}{8} + \epsilon \Delta \rho_{\ket{000}}
\end{eqnarray}
which is isomorphic to the pure state  $\ket{000}$ \cite{Corynmr_1st}.
In the pseudopure state, the traceless deviation part has the form 
\begin{eqnarray}
\Delta\rho_{\ket{000}} = && \frac{1}{4}(I_{1z}+I_{2z}+I_{3z} + 2I_{1z}I_{2z} \nonumber\\
&&+ 2I_{2z}I_{3z} + 2I_{1z}I_{3z} + 4I_{1z}I_{2z}I_{3z})
\label{Delta rho 000}.
\end{eqnarray}

The first spin, F$_1$, is used as an ancilla qubit, and other spins, F$_2$ and F$_3$, as the system qubits (see Fig. \ref{moussa}).  The initial Hadamard gate on the first spin prepares $\rho_\mathrm{\ket{+00}}$.  To measure $\expec{AB}_{\ket{00}}$, 
we apply the corresponding controlled operations $A$ and $B$ as indicated in circuit \ref{moussa}.  The transverse magnetization of the ancilla qubit will be
proportional to the expectation value $\expec{AB}_{\ket{00}}$.
The absolute value of $\expec{AB}_{\ket{00}}$ is estimated by normalizing
the value obtained in the above experiment with that obtained from a reference experiment having no controlled operations.  Similarly we can measure the
other expectation values $\expec{BC}_{\ket{00}}$, $\expec{CD}_{\ket{00}}$,
and $\expec{AD}_{\ket{00}}$, and determine the value of ${\bf I}_0$.
Other values ${\bf I}_l$ are obtained by preparing the corresponding
pseudopure states $\rho_{\ket{+01}}$, $\rho_{\ket{+10}}$, and $\rho_{\ket{+11}}$
and applying the circuit \ref{moussa} in each case.  

In our experiments, all the controlled operations were realized by numerically optimized radio frequency (RF) pulses obtained 
using GRAPE technique \cite{Khaneja}.  
Each pair of controlled operations in circuit \ref{moussa} was realized by
a GRAPE sequence with a duration of about 23 ms (having RF segments of duration  $5~\upmu s$) and an average Hilbert-Schmidt fidelity better than 0.99 over 10\% variation in RF amplitude.

\begin{figure*}
\includegraphics[width=12cm,clip = true,trim = 0cm 1cm 0cm 2cm]{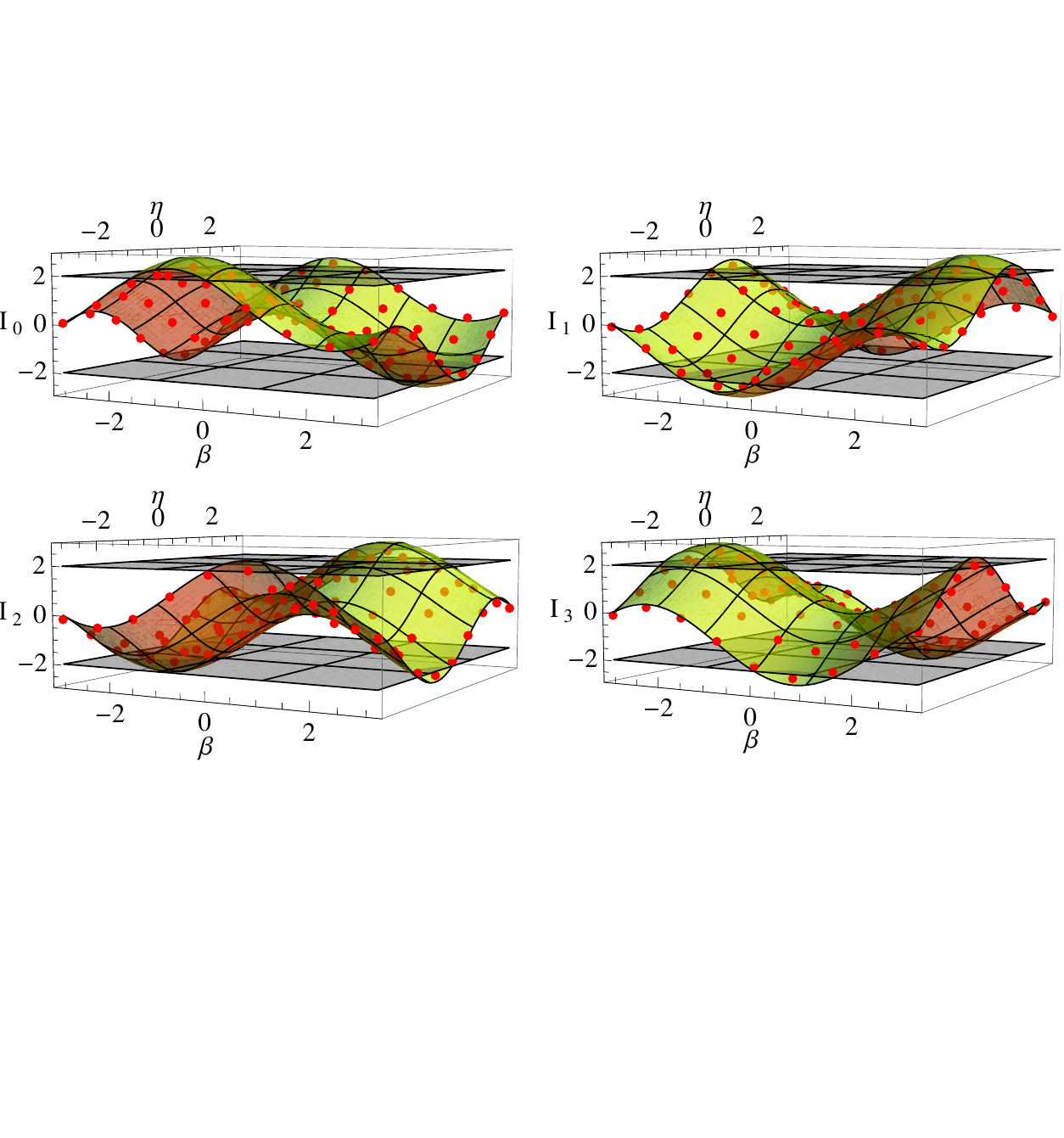}
\vspace*{-3.4cm}
\caption{$ \mathrm{\textbf{I}}_0 $,$ \mathrm{\textbf{I}}_1 $,$\mathrm{\textbf{I}}_2 $,and $ \mathrm{\textbf{I}}_3 $ represents evaluation of expression \ref{I_main} for eigenstates $ \ket{0}_{QHO},\ket{1}_{QHO},\ket{2}_{QHO}$, and $\ket{3}_{QHO} $
respectively. The curved surface represents theoretical values, and the points are experimental values. Flat planes at $ 2 $ and $ -2 $ represents classical bound.}
\label{results}
\end{figure*}

We estimated the values for $\textbf{I}_l$ (\ref{I_main}), for all the four eigenstates and 
independently varied both $\beta$ and $\eta$ over the range $[-\pi,\pi]$ with 
increments of $\pi/4$. The results are shown in Fig. \ref{results}.
The maximum theoretical violation is $ 2\sqrt{2} = 2.82$.
The experimental values of maximum violations for $\textbf{I}_0$, $\textbf{I}_1$, $\textbf{I}_2$, and $\textbf{I}_3$ are $ 2.40 \pm 0.02,~ 2.45 \pm 0.02,~ 2.39 \pm 0.02$, and $2.42 \pm 0.03$ respectively.

\subsection{State independent contextuality}
Hong-Yi Su \textit{et.al.} also studied the state independent contextuality \cite{Cont_theory,Cabello_state_ind_context}.
They considered the inequality (arising from NCHV model)
\begin{eqnarray}
&&\expec{P_{11}P_{12}P_{13}} + \expec{P_{21}P_{22}P_{23}} + \expec{P_{31}P_{32}P_{33}} \nonumber  \\
&&+\expec{P_{11}P_{21}P_{31}}+\expec{P_{12}P_{22}P_{32}}-\expec{P_{13}P_{23}P_{33}} \le 4~~~~~~
\label{stateIn}
\end{eqnarray}
where $ P_{ij} $ are the elements of the matrix P,
\begin{equation}
P=\left(
\begin{array}{ccc}
\Gamma_z & \Gamma_z' & \Gamma_z\Gamma_z' \\
\Gamma_x' & \Gamma_x & \Gamma_x\Gamma_x'\\
\Gamma_z \Gamma_x' & \Gamma_x \Gamma_z' & \Gamma_y\Gamma_y'
\end{array}
\right).
\end{equation}

Operators in each row of the matrix $P$ commute with each other. Similarly in each column. $P_{ij}$ are dichotomic observables with measurement outcomes $\pm 1$. 
We can verify the inequality \ref{stateIn} by preassigning the values $\pm 1$ 
to each observable $P_{ij}$.

Now introducing the operators from expressions (\ref{Gammas defined}), we find
that the product of each row of the matrix $P$ is identity (i.e., $ P_{j1}P_{j2}P_{j3} = \mathbbm{1}$) having eigenvalue $+1$.  Similarly, the products along each of the first two columns is again identity.  However, the product along the last column, i.e., $P_{13}P_{23}P_{33} = -\mathbbm{1}$ having the eigenvalue $-1$.  No preassignment of $\pm 1$ to the various elements of $P$ can satisfy the
condition that, product along each row and along the first two columns be $+1$ and
along the last column be $-1$.  This shows that quantum theory is not compatible with NCHV model.  
Further, for the above choice of operators, the expectation values for the first five operators in expression 
\ref{stateIn} are all $+1$ while that of the last term is $-1$.  Therefore, quantum bound for lhs of expression \ref{stateIn} is $6$ independent of initial state of the system.

To investigate  state-independent QC, we need to measure joint expectation values of three observables.  We again use the circuit \ref{moussa} for this purpose.
Taking advantage of the state independent property of the above mentioned inequality, we choose thermal equilibrium state (\ref{rho_eq}) as initial state. A $(\pi/2)_y $ pulse was applied on the first spin to prepare the ancilla
in a superposition state. Then state (\ref{rho_eq}) takes the form: $ (1-4\epsilon) \mathbbm{1}_8/8 + \epsilon (\outpr{+}{+}\otimes\mathbbm{1}\otimes\mathbbm{1})+\epsilon(I_{2z}+I_{3z})$.

All the controlled $P_{ij}$ operations were realized using the GRAPE sequences having average fidelities better than $ 0.99 $ over 10\% variation in RF amplitude.
The total duration of the RF sequences (for each term in \ref{stateIn}) were about 40 ms.
Experimentally obtained value of lhs of inequality \ref{stateIn} is $ 4.81 \pm 0.02$.

Thus we observed a clear violation of the classical bound. However it is still lower than the quantum limit.  The reduced violation is attributed to decoherence ($ \mathrm{T}_2 $ decay) and imperfections in RF pulses. 

\section{Conclusion}
We have experimentally demonstrated the quantum contextuality exhibited by first four energy eigenstates of a one dimensional quantum harmonic oscillator through the violation of an inequality obtained from a noncontextual hidden variable model.
The continuous observables of the harmonic oscillator are mapped onto pseudospin observables which are then experimentally realized on a pair of qubits.
We used Moussa protocol to retrieve the joint expectation values of observables using an ancillary qubit.  
Our quantum register was based on three mutually interacting spin-1/2 nuclei controlled by NMR techniques.
We also demonstrated a violation of an inequality formulated to study state-independent contextuality, by measuring a set of expectation values on
the thermal equilibrium states of the nuclear spins.
The results of the experiment not only establish the validity of quantum theoretical
calculations, but also highlights the success of NMR systems as quantum simulators.

\section*{Acknowledgement}
The authors are grateful to Prof. Dipankar Home for referring us to Hong-Yi Su \textit{et.al}'s paper \cite{Cont_theory} and also for his valuable suggestions and discussions. We also thank Siddharth and Abhishek Shukla for useful discussions.

\bibliographystyle{apsrev4-1}
\bibliography{bib_ch}

\end{document}